\documentclass[fleqn,usenatbib]{mnras}

\usepackage[T1]{fontenc}
\usepackage{ae,aecompl}

\usepackage{graphicx}	
\usepackage{amsmath}	
\usepackage{amssymb}	
\usepackage{pdflscape}
\usepackage{times,txfonts}
\usepackage{color}
\usepackage{hyperref}	

\definecolor{linkcolor}{rgb}{0,0,0.25}
\hypersetup{
  colorlinks=true,        
  linkcolor=linkcolor,    
  citecolor=linkcolor,    
  filecolor=linkcolor,    
  urlcolor=linkcolor      
}

\usepackage{multirow}
\usepackage{rotating}
\usepackage{lscape}
\usepackage{tablefootnote}
\usepackage{hyperref}
\usepackage{listings}

\usepackage[colorinlistoftodos]{todonotes}

\definecolor{salmon}{rgb}{0.95,0.5,0.25}

\newcommand{\gcmass}{9.2}
\newcommand{\gcemass}{0.4}

\newcommand{\gcrad}{3.7}
\newcommand{\gcerad}{0.1}

\newcommand{\gaia}{\emph{Gaia}}

\title[Made-to-Measure Modelling of Globular Clusters]{Made-to-Measure Modelling of Globular Clusters}
\author[Webb, Hunt, \& Bovy]{Jeremy J. Webb$^1$\thanks{E-mail: webb@astro.utoronto.ca (JW)}, Jason A. S. Hunt$^2$ \& Jo Bovy$^1$
 \\
$^1$David A. Dunlap Department of Astronomy and Astrophysics, University of Toronto, 50 St. George Street, Toronto, ON, M5S 3H4, Canada \\
$^2$Center for Computational Astrophysics, Flatiron Institute, 162 5th Avenue, New York, NY 10010, USA
}

\begin{document}

\pagerange{\pageref{firstpage}--\pageref{lastpage}} \pubyear{2021}

\maketitle

\label{firstpage}

\begin{abstract}
We present the first application of the made-to-measure method for modelling dynamical systems to globular clusters. Through the made-to-measure algorithm, the masses of individual particles within a model cluster are adjusted while the system evolves forward in time via a gravitational $N$-body code until the model cluster is able to reproduce select properties of an observed cluster. The method is first applied to observations of mock isotropic and anisotropic clusters while fitting against the cluster's three dimensional or projected density profile, density weighted mean-squared velocity profile, or its density profile with individual mean-squared velocity profiles. We find that a cluster's three-dimensional density profile can easily be reproduced by the made-to-measure method, with minor discrepancies in the outer regions if fitting against a cluster's projected surface density or projected kinematic properties. If an observed cluster is anisotropic, only fitting against the cluster's density profile and individual mean-squared velocity profiles will fully recover the full degree of anisotropy. Partial anisotropy can be recovered as long as two kinematic properties are included in the fit. We further apply the method to observations of the Galactic globular cluster M4 and generate a complete six-dimensional representation of the cluster that reproduces observations of its surface density profile, mean-squared proper motion velocity profile, and mean-squared line of sight velocity profile. The M2M method predicts M4 is primarily isotropic with a mass of $\gcmass \pm \gcemass \times 10^4\, M_{\odot}$ and a half-mass radius of $\gcrad\pm\gcerad$ pc.

\end{abstract}


\section{Introduction} \label{s_intro}

Stars do not form in isolation, but in clustered environments that in some cases can be gravitationally bound \citep{Lada03, Kruijssen12}. The first star clusters to form in a galaxy, commonly referred to as globular clusters, do so at high redshifts while galaxies are being assembled \citep{Kruijssen15, Forbes18}. Once formed, the evolution of a globular cluster is driven by stellar evolution \citep{Hurley2000, Hurley2002}, two-body relaxation \citep{Heggie03}, and tidal stripping \citep{Baumgardt03}. The latter mechanism has been shown to be quite sensitive to the distribution of matter within the cluster's host galaxy \citep[e.g][]{Gieles07, Kruijssen2009a, Webb14, Webb2015, Webb20}. Knowledge of an observed cluster's dynamical state, which is the combined distribution of stellar masses, positions, and velocities, allows for a quantification of how a cluster has changed since birth assuming an initial distribution function for newborn clusters is known. Understanding how a cluster has evolved since formation opens the door to use clusters as tools to study both star formation and the evolution of the galaxy within which the cluster resides.

An extensive list of distribution functions, both observationally and dynamically motivated, have been applied to observations of star clusters. The most common isotropic models include \citet{Plummer1911}, \citet{Woolley1954}, \citet{King1962}, \citet{Michie1963}, \citet{Sersic1963}, \citet{King1966}, \citet{Wilson1975}, \citet{Gunn1979}, \citet{Elson1987} and \citet{Bertin2008}. The \citet{Woolley1954}, \citet{King1966}, and \citet{Wilson1975} models are part of a larger family of lowered isothermal models, which can be modified to include orbital anisotropy \citep{Gieles2015}. Additional variations of the traditional dynamical models include the effects of rotation \citep{Varri2012} and the presence of potential escapees \citep{Claydon2019}. 

Historically, when applying one of the above models to observations, the observational data is radially binned to create a projected surface brightness profile, number density profile, or mass density profile. A given model's parameters are then sampled until the best fit model is found, often using $\chi^2$ as a metric for goodness of fit \citep[e.g.][]{McLaughlin05}. More recent studies have been able to include a cluster's kinematic properties when comparing observations to dynamical models thanks to surveys like \gaia\ \citep{Gaia16a,gaia18}, the Hubble Space Telescope Proper Motion (HSTPROMO) Survey \citep{Bellini2014} and the Apache Point Observatory Galactic Evolution Experiment \citep[APOGEE;][]{APOGEE2020}. The kinematic properties of stars can be used to confirm  membership before measuring a cluster's density profile \citep{deboer18} or can be used directly to measure a cluster's velocity dispersion profile along a given direction \citep[e.g.][]{Jindal2019}. In the latter case, it becomes possible to simultaneously fit a cluster's density profile and kinematic profile with a dynamical model \citep[e.g.][]{Zocchi2017}. However when fitting against a cluster's density profile and kinematic profile via a $\chi^2$ method, it must be decided whether more or less weight should be attributed to the individual components as their uncertainties and spatial coverage will likely differ. When including spatial and kinematic data, Bayesian inference offers an improvement over the $\chi^2$ method of fitting dynamical models to observations \citep{Eadie22}.

The main disadvantage of fitting dynamical models to observations of star clusters is that a dynamical model has to be assumed in the first place. While other factors such as binning, fitting technique, and completeness will also affect how well a model fits observations, whether or not the functional form of a given model's distribution function is an accurate representation of a star cluster will always be the limiting factor \citep{HenaultBrunet2019}. Alternative methods that do not require an assumed distribution function include Jeans modelling \citep{Sollima2015} or the  direct comparison of observational data to N-body simulations \citep{Baumgardt2017a, Baumgardt2018}. Jeans modelling is less effective in the outer regions of a star cluster where the density is low, which can result in large uncertainties in estimates of a cluster's mass or half-mass radius \citep{HenaultBrunet2019}. Comparing observations to $N$-body simulations requires both a large suite of simulations and interpolation between simulations. Given the computational expense of direct $N$-body simulations of over 100,000 stars, it is extremely difficult to properly model massive clusters since the full parameter space of initial cluster properties (mass, size, stellar mass function, orbit, density profile) cannot be realistically covered.
    
Made-to-measure modelling \citep[M2M;][]{Syer96} offers another alternative method for determining the dynamical state of a globular cluster that doesn't require a large suite of simulations or an assumed functional form of the underlying distribution function. M2M is a flexible algorithm for tailoring a particle based model to match some desired target state, such as a distribution function, another particle based model or real observational data. While the traditional implementation of \cite{Syer96} fits the observables by weighting a set of tracer particles \citep[similar to the better known Schwarzchild method which weights orbits;][]{Schwarzschild}, it has also been adapted to tailor self-gravitating systems \citep{DegPhD,Hunt2013a} by using the algorithm to directly modify the particles' masses.

To date, M2M has been used to model several galactic scale systems, both external \citep[e.g.][]{deLorenzi07,deLorenzi08,Long10,LM12,Das11,Zhu14,Williams15} and the Milky Way \citep[e.g.][]{Bissantz04,LM13,Portail15,Bovy18}. It has also been used to set up stable initial conditions for $N$-body simulations \citep{Dehnen09,Malvido15}.


    
In this study, we present the first ever application of the M2M method to star clusters. In Section \ref{s_method} we summarize the method and discuss how it has been implemented in the \texttt{m2mcluster}\footnote{https://github.com/webbjj/m2mcluster} python package with the help of the Astrophysical Multipurpose Software Environment \citep[\texttt{AMUSE};][]{PortegiesZwart2018}. As a proof of concept, we first apply the method to isotropic and anisotropic mock clusters in Section \ref{s_results}. We then apply the M2M method to the Galactic globular cluster M4. Finally we discuss and summarize our findings in Sections \ref{s_discussion} and \ref{s_conclusion}.

\section{Method} \label{s_method}

The M2M method was first introduced by \citet{Syer96} and has been subject to several modifications over time, including the addition of kinematic constraints \citep{deLorenzi07,deLorenzi08} and the ability to model individual particle data taking into account individual errors instead of binned distributions \citep{Hunt2013a,Hunt2014}. In the following sections we summarize the M2M formalism adopted in this study and discuss how it has been implemented in the \texttt{m2mcluster} python package.

\subsection{Made-to-Measure}

In the traditional M2M implementation, the observables of the system we are intending to reproduce are given by
\begin{equation}
Y_j=\int K_j(z)f(z)d^6z,
\end{equation}
where $j$ represents each individual observable, $z=(\mathbf{r},\mathbf{v})$ are the phase space coordinates, $f(z)$ is the distribution function of the target system, and $K_j$ is a known kernel which controls the contribution of neighboring data points to each observable. These observables are most commonly some form of density or kinematics. The equivalent model observable is given by 
\begin{equation}
y_j=\sum^N_{i=1}w_iK_j[\mathbf{z}_i(t)],
\end{equation}
where $w_i$ are the particle weights, and $\mathbf{z}_i$ are the phase space coordinates of the $i^{th}$ model particle. We then calculate the difference between each corresponding model and target observable simply as
\begin{equation}
\Delta_j=\frac{y_j(t)-Y_j}{Y_j}.
\end{equation}
This $\Delta_j$ is then used to calculate the magnitude of the change in the particle weights per timestep as influenced by the $j^{th}$ observable
\begin{equation}
\frac{\mathrm{d}}{\mathrm{d}t}w_i(t)=-\epsilon w_i(t)\sum_j\frac{K_j[\mathbf{z}_i(t)]}{Z_j}\Delta_{y_j}(t),
\label{dwdt1}
\end{equation}
where $\epsilon$ controls the rate of change and $Z_j$ is a so far arbitrary constant for normalisation. In practice, the ratio $K_{ij}/Z_j$ will control how much each particle contributes to each observable. The system is then evolve forwards in time via a gravitational dynamics code, with particle weights changing as per Equation \ref{dwdt1}.

Since in most applications we will have more particles than observables, $N>>j$, equation (\ref{dwdt1}) is ill conditioned. This issue can be avoided by introducing the entropy term from \cite{Syer96}; see within for the derivation. This addition changes equation \ref{dwdt1} to 
\begin{equation}
\frac{\mathrm{d}}{\mathrm{d}t}w_i(t)=-\epsilon w_i(t)\biggl[\sum_j\frac{K_j[\mathbf{z}_i(t)]}{Z_j}\Delta_j(t)+\mu\ln\biggl(\frac{w_i(t)}{\hat{w}_i}\biggr)\biggr],
\label{dwdt2}
\end{equation}
where $\mu$ is a parameter to control the strength of this regularisation and $\hat{w}_i$ is some prior set of weights, commonly taken to equal the initial particle weights \citep[although see][for a more sophisticated regularisation scheme]{Morganti12}.

To incorporate an observational measurement error $\sigma_j$ into the algorithm, as per \citet{Bovy18}, it is helpful to define the term $\chi^2$, which can be written as:

\begin{equation}
    \chi_j^2=\frac{(y_j(t)-Y_j)^2}{\sigma_j^2}
\end{equation}

Defining $\chi^2$ in this way allows us for Equation \ref{dwdt2} to be written as:
 
 \begin{equation}
\frac{\mathrm{d}}{\mathrm{d}t}w_i(t)=-\epsilon w_i(t)\biggl[\frac{1}{2}\sum_i \frac{d\chi_j^2}{d w_i}+\mu\ln\biggl(\frac{w_i(t)}{\hat{w}_i}\biggr)\biggr],
\label{dwdtf}
\end{equation}

where $Z_j$ has been folded into the kernel being used to measure the observable parameter.  In this formalism, \citet{Bovy18} defines $\Delta_j$ as simply $(y_j(t)-Y_j)$. In cases where there are multiple observables, additional $\frac{1}{2}\sum \frac{d\chi_j^2}{d w_i}$ terms can be added to equation \ref{dwdtf}. 
 
We also make use of an extension to the M2M algorithm that allows for a smoothing of the $\Delta_j$ term between timesteps. As per \citet{Syer96} and \citet{Bovy18}, this smoothing is governed by the $\alpha$ parameter via:

\begin{equation}
    \frac{d \tilde{\Delta}_j}{dt}=\alpha(\Delta_j-\tilde{\Delta}_j)
\end{equation}

where $\tilde{\Delta}_j$ then replaces $\Delta_j$ in Equation \ref{dwdtf}. This step lessens the contribution of Poisson noise due to the fixed number of particles in the model \citep{Syer96}.

\subsection{Implementation}

In order to apply the M2M method to globular clusters, we have developed the \texttt{m2mcluster} python package. In addition to the M2M algorithm itself, \texttt{m2mcluster} makes use of AMUSE to model the gravitational $N$-body evolution of the system between time steps. The key benefits to using AMUSE to model the system's gravitational evolution is that several different options exist to model the cluster's evolution, particle masses can be extracted and updated at every time step, particles can be easily added and removed from the simulation, and the simulation time step duration can be easily updated while the simulation is running. These features allow for particle weights to be changed at every time step as determined by the M2M algorithm, massive particles can be split into several low-mass particles, particles with weights that move below our minimum threshold can be removed from the simulation, and the time step size can be altered if the system's dynamical time changes dramatically.

While any of the community stellar dynamics codes within AMUSE can be used to model cluster evolution, in this study we specifically use BHTree \citep{Barnes1986} with an opening angle of 0.6 and gravitational softening length of 0.01 pc. The minimum and maximum particle weights in our simulation are set to 0.1 $M_{\odot}$ and 2.0 $M_{\odot}$ respectively, such that the initial weights of 1.0 $M_{\odot}$ are roughly in the centre of the allowable range. If a particle's weight is set to below the minimum mass it is removed from the simulation. If a particle's weight is set to above the maximum mass it is replaced by N 1.0 $M_{\odot}$ particles, where the choice of N ensures the total mass of the initial particle is recovered. The new particles are all given the same clustercentric radius and velocity as the particle that they are replacing, but their $\phi$ and $\theta$ coordinates are randomized.

Currently, no external tidal field is included in the simulation. However, it is possible to set the maximum distance a particle can reach before being removed from the simulation. For comparisons to mock clusters or observations, we set the maximum distance to either the maximum stellar distance in the observed dataset or the cluster's tidal radius. 

\section{Results} \label{s_results}

To implement and validate the application of the M2M method to star clusters, we apply the method to mock clusters for which we know each star's mass, position and velocity. To test the strength of the method we apply it to both an isotropic star cluster and an anisotropic cluster. Finally we apply the method to observations of the Galactic globular cluster M4. In all cases, we calculate observable parameters assuming a normalized Gaussian kernel $K_j(r)$ (or $K_j(R)$ in projection) centred on the the radial bin that a given star particle resides. For data with logarithmically spaced radial bins, the kernel is $K_j(\log r)$ or $K_j(\log R)$. The dispersion of the Gaussian is set equal to half the bin's width \citep{Syer96}. The normalization is such that the total contribution of a given particle to all bins sums to 1, and then the contribution to each bin is further normalized by the volume of the bin (or area when working with projected observables).

\subsection{Isotropic Star Clusters}\label{s_results_iso}

We begin by first trying to reproduce the three-dimensional density profile of a mock cluster using the M2M method, where the model density is given by $\nu(r_j)=\sum_i w_i K_j(r_i)$. The mock cluster is a \citet{King1966} model with $\Phi_0 = 1$, a virial radius of 10 pc, and a mass of 10,000 $M_{\odot}$. For the initial model cluster, we purposefully choose a model that has a dynamical state very far from the mock cluster's dynamical state. More specifically we chose a \citet{King1966} model with $\Phi_0 = 10$, a virial radius of 5 pc, and a mass of 12,000 $M_{\odot}$. For both the mock and model cluster, stellar positions and velocities are sampled with LIMEPY \citep{Gieles2015}. All stellar masses are set to 1 $M_{\odot}$. The mock cluster's density profile is determined by generating 20 equally spaced bins between the cluster's centre and the outermost star. The model cluster's density profile is always measured within the same bins. The uncertainties in any measured profiles is always assumed to be $10\%$. When running the M2M algorithm we impose a maximum distance of 30 pc, which corresponds to the maximum stellar distance in the mock cluster.

\begin{figure*}
    \includegraphics[width=\textwidth]{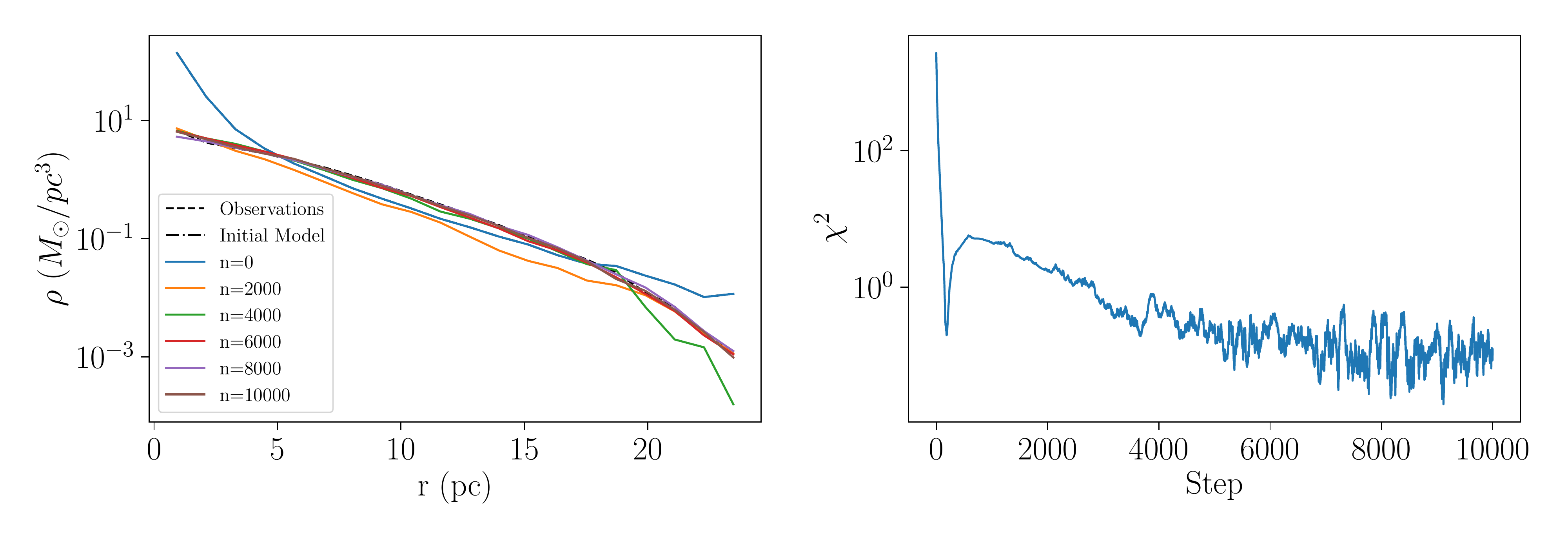}
    \caption{Left Panel: Evolution of the model cluster's density profile between the first and last step (step numbers are marked in the legend as n). For comparison purposes, the density profile of the target isotropic mock cluster is also illustrated. Right Panel: The $\chi^2$ difference between the model and isotropic mock cluster's density profiles as a function of step. Over 10,000 steps, the model cluster's density profile converges to that of the mock cluster.}
    \label{fig:rhofit_evol}
\end{figure*}

In our first application of the M2M method to a mock cluster, we only compare the density profiles of the mock and model clusters. The left panel of Figure \ref{fig:rhofit_evol} illustrates the change in the model cluster's density profile as the M2M method adjusts particle masses while the cluster evolves dynamically. Over 10,000 steps, the model cluster converges and is able to reproduce the observed density profile exactly. The right panel of Figure \ref{fig:rhofit_evol} further demonstrates how $\chi^2$ is reduced with each step until it converges to a constant value, where $\chi^2 = \sum_i^N(\rho_{i,model}-\rho_{i,observations})^2/\rho_{i,observations})$  Through a moderate exploration of the parameter space, we find $\epsilon=0.005$ and $\mu=0.005$ are the ideal values for the M2M parameters given this mock cluster. We did not assume any smoothing between time-steps ($\alpha=0$). The size of each time step between applications of the M2M algorithm is adaptive, and set equal to 0.1 $\%$ of the cluster's instantaneous dynamical time. Within AMUSE, the dynamical time is the free-fall timescale of a pressureless homogeneous sphere that has the same size and average density as the cluster \citep{PortegiesZwart2018}. The duration of the very first time step is 0.06 Myr.

With the M2M method successfully reproducing the density profile of a mock cluster when fitting against the density profile itself ($\rho$ fit), we next extend the fitting process to include other cluster parameters. Moving forward, we also fit against the mock cluster's density weighted mean velocity profile ($\rho v^2$ Fit), the mock cluster's density profile with all three independent velocity components ($\rho$,$v_r$,$v_{\phi}$,$v_{\theta}$ \ Fit), the mock cluster's projected density profile ($\Sigma$ Fit), the mock cluster's projected density weighted mean line-of-sight velocity profile ($\Sigma v_{los}^2$ Fit), and the mock cluster's projected density profile with two projected velocity components ($\Sigma$,$v_R$,$v_T$ Fit). The density weighted mean velocity is calculated via $\mathrm \nu<v^2>(r_j)=\sum_i w_i v_i^2 K_j(r_i)$, while the mean square velocity profiles are each calculated via $\mathrm <v^2>(r_j)=\sum_i w_i v_i^2 K_j(r_i) / \sum_i w_i K_j(r_i)$ \citep{Bovy18}.

\begin{figure}
    \includegraphics[width=0.48\textwidth]{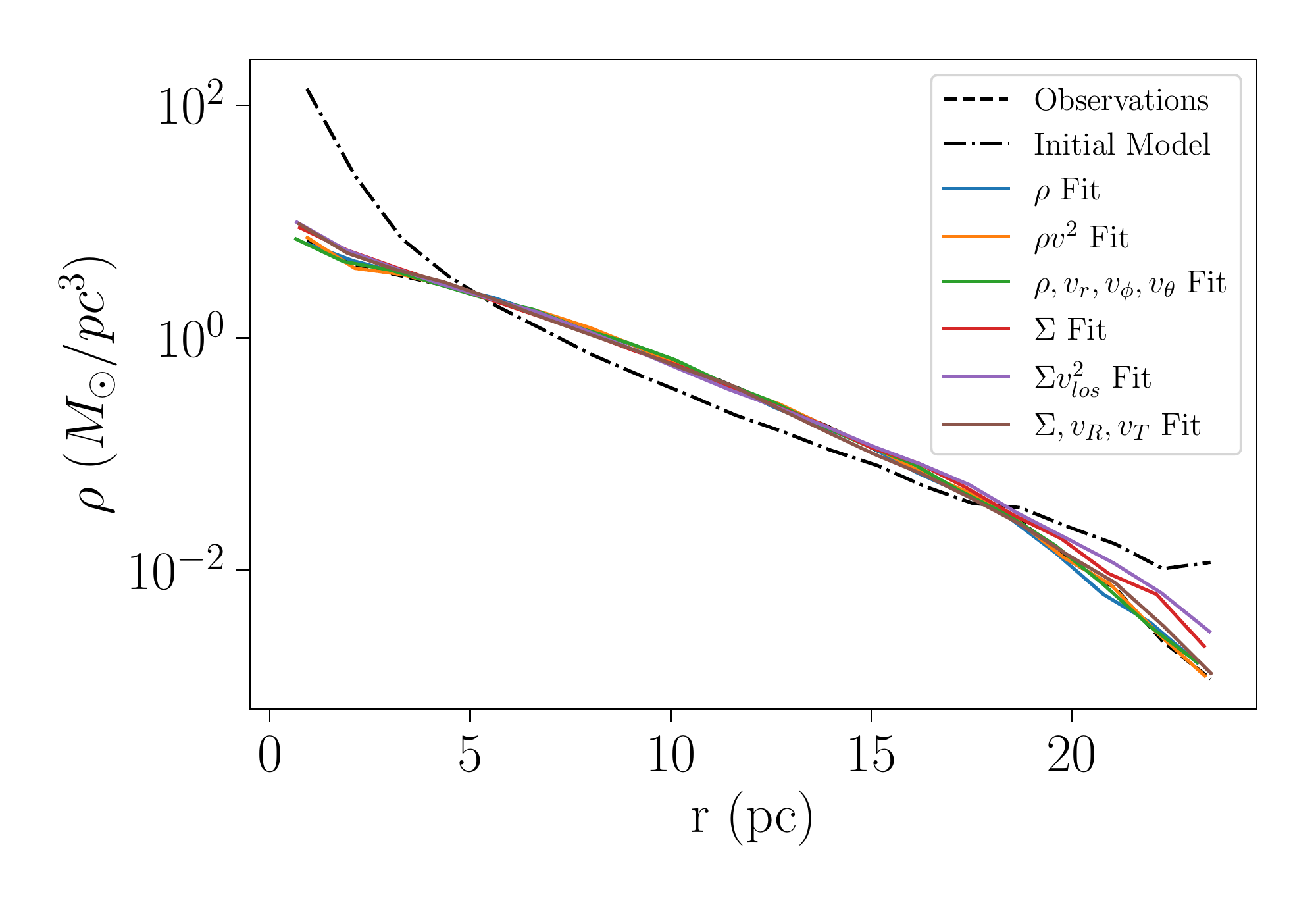}
    \caption{Density profile of model clusters fit to an isotropic mock cluster using different structural and kinematic parameters, with the fitting parameters marked in the legend. For comparison purposes, the density profile of the initial model cluster (dash dot line) and mock cluster (dashed line) are also illustrated. In all cases, the model cluster's density profile converges to that of the mock cluster.}
    \label{fig:rhofit}
\end{figure}

Figure \ref{fig:rhofit} illustrates the final density profile of model clusters that have been fit to the mock cluster using the M2M method. In all cases, the mock cluster's three dimensional density profile is recovered with $\chi^2$ values less than 0.6. Fitting against $\rho$ only expectantly provides the best fit as there are no other constraints placed on the cluster as it evolves. Including kinematics in the fitting procedure only slightly increases the $\chi^2$ between the mock cluster and model cluster's density profiles. When only fitting against projected properties of the mock cluster, the three dimensional density profile is still well recovered out to approximately 20 pc. It is only in the outermost regions of the cluster that fitting against the mock cluster's projected density (with or without kinematic information) results in a slightly poorer fit to the three dimensional density profile.

\begin{figure*}
    \includegraphics[width=\textwidth]{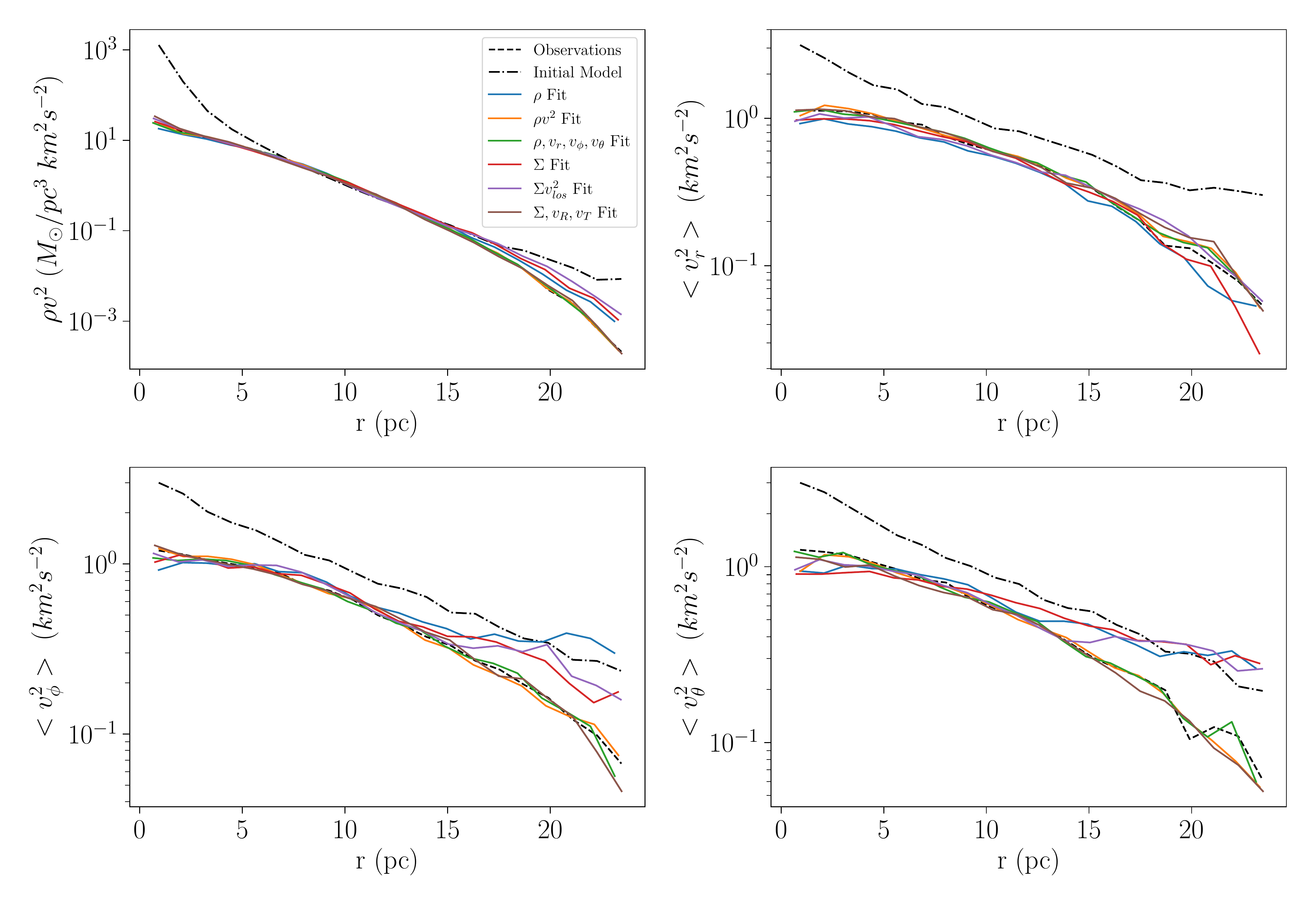}
    \caption{Density weighted mean-squared velocity profile (top-left panel), mean-squared radial velocity profile (top-right panel), mean-squared azimuthal angle velocity profile (bottom-left panel), and mean-squared polar angle velocity profile (bottom-right panel) of model clusters fit to isotropic mock clusters using different structural and kinematic parameters, with the fitting parameters marked in the legend. For comparison purposes, the corresponding initial model cluster profile (dash dot line) and mock cluster profile (dashed line) are also illustrated. In all cases, the model cluster's inner region profiles converge to those of the mock cluster. However, only when fitting against the mock cluster's density profile \textit{and} all three velocity profiles are the kinematics of the mock cluster's outer regions reproduced by the model cluster.}
    \label{fig:kinfit}
\end{figure*}

With the mock cluster's density profile well reproduced by each set of fitting parameters, we now consider how well other properties of the cluster are reproduced by the M2M method. Figure \ref{fig:kinfit} compares the final kinematic profiles of model clusters to those of the mock cluster. When fitting to the density profile only, both in three-dimensions and in projection, the cluster's central kinematics are well reproduced. However the cluster's mean-squared velocity profiles in the outer regions are all overestimated. Hence, with only spatial information, it is possible to generate a model cluster that is in equilibrium, reproduces an observed cluster's density profile, but doesn't necessarily reproduce the observed cluster's kinematics.

Including one or more kinematic properties in the M2M method significantly increases the M2M algorithm's ability to generate a true representation of an observed cluster. However, fitting against the cluster's density weighted mean-squared velocity profile still slighty overestimates the mean-squared azimuthal velocity in the outermost regions of the cluster. Only fitting against the three dimensional density profile and all three individual kinematic profile results in the model cluster and mock cluster being in complete agreement.

\begin{figure*}
    \includegraphics[width=\textwidth]{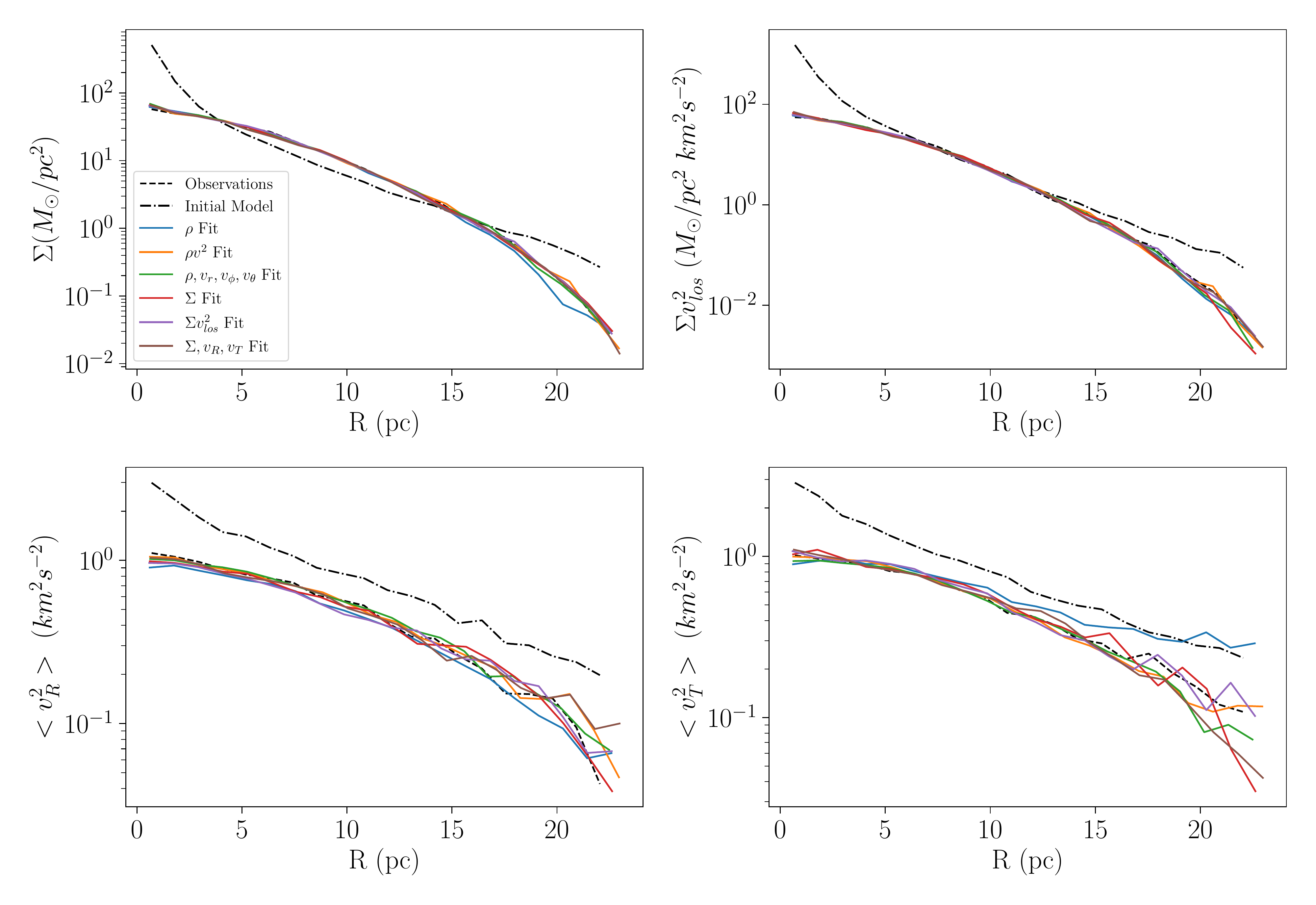}
    \caption{Surface density profile (top-left panel), density weighted mean-squared line-of-sight velocity profile (top-right panel), mean-squared radial velocity profile (bottom-left panel), mean-squared tangential velocity profile (bottom-right panel) of model clusters fit to isotropic mock clusters using different structural and kinematic parameters, with the fitting parameters marked in the legend. For comparison purposes, the corresponding initial model cluster profile (dash dot line) and mock cluster profile (dashed line) are also illustrated. In all cases, the model cluster's inner region profiles converge to those of the mock cluster. However, fitting against just the cluster's density or surface density profiles results in the model clusters having difficulty reproducing the projected kinematics of the mock cluster's outer regions.}
    \label{fig:sigfit}
\end{figure*}

Fitting against the cluster's surface density profile and projected kinematics typically results in the model cluster overestimating three-dimensional mean-squared velocities in the outer regions of the cluster. However as Figure \ref{fig:sigfit} illustrates, the final projected kinematic profiles of model clusters that have been fit to an observed dataset using the M2M method are well reproduced. In fact, whether fitting against three-dimensional or projected data, the M2M method accurately reproduces the projected outer regions of the observed mock cluster, with a minor discrepancy when fitting against the cluster's three-dimensional density profile only. The fact that fitting against a mock cluster's projected properties doesn't necessarily reproduce the mock cluster's three dimensional properties highlights how information is lost when viewing clusters in projection.

\subsection{Anisotropic Star Clusters}

To further test the robustness of the M2M method and its application to star clusters, we aim to reproduce a mock cluster with radial anisotropy. To initialize a mock cluster with radial anisotropy, we use the same \citet{King1966} model parameters as before with the anisotropy radius was set equal to 0.3 pc. The anisotropy radius sets the degree of orbital anisotropy in the cluster, with the cluster's inner regions remaining isotropic and the degree of radial anisotropy in the outer regions increasing with decreasing $r_a$ \citep{Michie1963, Gieles2015}. The initial model cluster was given the same model parameters as the mock cluster, but with no radial anisotropy. Similar to the isotropic case, initial mock cluster and model cluster positions and velocities were sampled using LIMEPY \citep{Gieles2015}, stellar masses were taken to be 1 $M_{\odot}$, and uncertainties in any measured profiles are taken to be $10\%$. The imposed maximum radius for the anisotropic test is 38.8 pc pc, which corresponds to the maximum stellar radius in the anisotropic mock cluster. It should also be noted that for the $\rho$,$v_r$,$v_{\phi}$,$v_{\theta}$ \ Fit case $\epsilon$ and $\mu$ had to be increased to 0.1, otherwise the change in particle weights was too slow and dynamical evolution eventually resulted in the cluster dissolving.

\begin{figure*}
    \includegraphics[width=\textwidth]{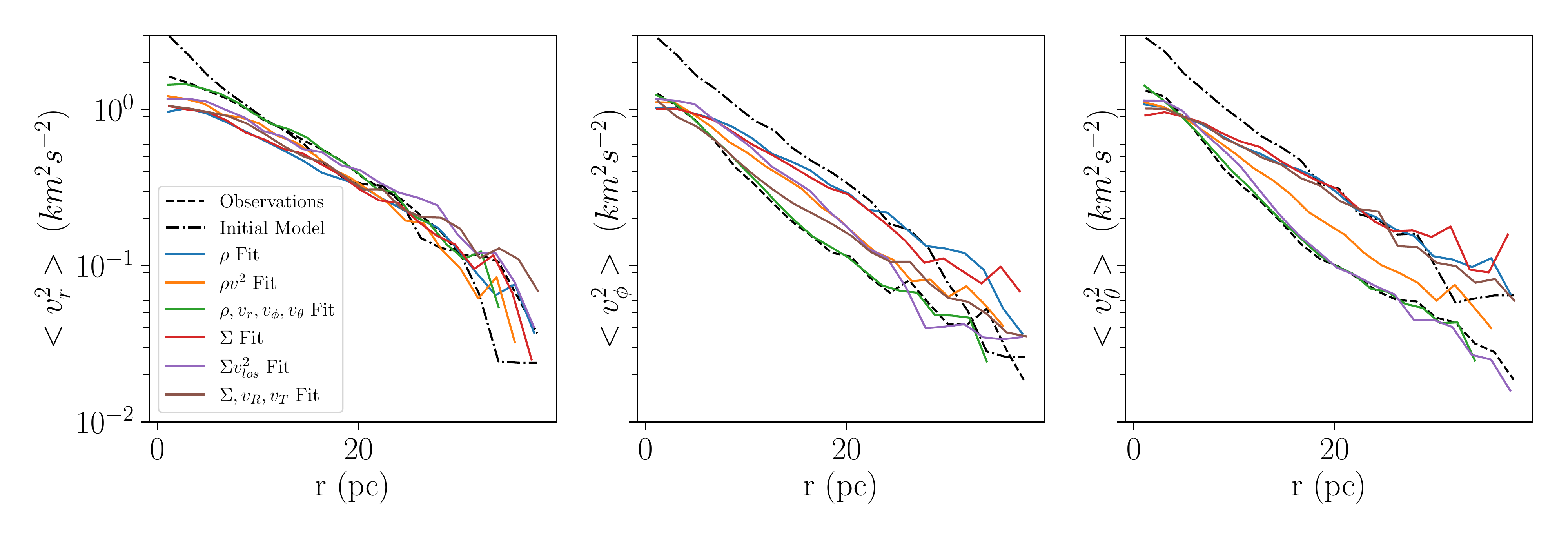}
    \caption{Mean-squared radial velocity profile (left panel), mean-squared azimuthal angle velocity profile (middle panel), and mean-squared polar angle velocity profile (right panel) of model clusters fit to an anisotropic mock cluster using different structural and kinematic parameters, with the fitting parameters marked in the legend. For comparison purposes, the corresponding initial model cluster profiles (dash dot line) and mock cluster profiles (dashed line) are also illustrated. Only when fitting against the mock cluster's density profile \textit{and} all three kinematic profiles is the radial anisotropy of the mock reproduced by the model cluster. }
    \label{fig:kinfit_aniso}
\end{figure*}

Similar to the isotropic cases, the three-dimensional and and projected density profiles of the mock cluster are easily reproduced by the M2M method. Figure \ref{fig:kinfit_aniso} demonstrates that completely recovering a cluster's radial anisotropy requires three-dimensional knowledge of the observed cluster's kinematics. Fitting only against the cluster's density profile or density weighted mean-squared velocity profile, either in three dimensions or projection, will always yield a near-isotropic model cluster since our initial model cluster is isotropic. Fitting against the cluster's density profile and three-dimensional kinematic profiles results in a model cluster with the same radial anisotropy profile as the observed mock cluster. Knowing only the clusters projected density profile, mean-squared projected radial velocity profile, and mean-squared projected tangential velocity profile partially reproduces the cluster's anisotropy profile, but overestimates the mean-squared azimuthal and polar velocity profiles.

\begin{figure*}
    \includegraphics[width=\textwidth]{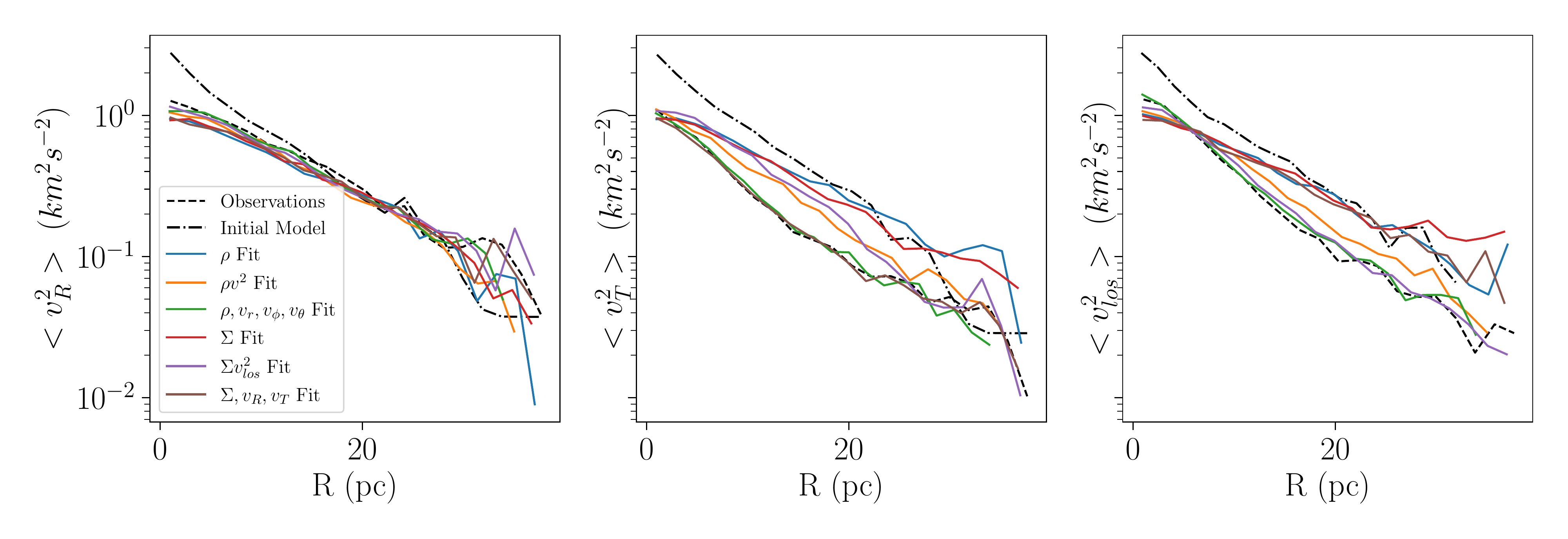}
    \caption{Mean-squared projected radial velocity profile (left panel), mean-squared projected tangential velocity profile (middle panel), and mean-squared line of sight velocity profile (right panel) of model clusters fit to an anisotropic mock cluster using different structural and kinematic parameters, with the fitting parameters marked in the legend. For comparison purposes, the corresponding initial model cluster profiles (dash dot line) and mock cluster profiles (dashed line) are also illustrated. Fitting against the mock cluster's projected density, radial velocity, and tangential velocity profiles partially reproduces the mock cluster's radial anisotropy.}
    \label{fig:kinfit_aniso_pro}
\end{figure*}

Figure \ref{fig:kinfit_aniso_pro} further demonstrates how the M2M method can reproduce an anisotropic cluster's mean-squared projected radial velocity profile and mean-squared projected tangential velocity profile when fitting against these parameter's. Unfortunately, without knowledge of kinematics along a third dimension, the model cluster will be less anisotropic than the observed mock cluster.

\subsection{Globular Cluster M4}

We have demonstrated in the previous two sections that the M2M method is capable of reproducing the density and kinematic profiles of isotropic and anisotropic mock clusters in three dimensions and in projection. The next step is to apply the method to an observational dataset. For the first application of the M2M method to star cluster observations, we have selected the closest Galactic globular cluster M4. 

Observational data for M4 is primarily taken from the \citet{Baumgardt2018} Galactic Globular Cluster catalogue and references therein. As per \citet{Baumgardt2017a}, the cluster's surface density profile was determined from the surface brightness profile measured by \citet{Trager95}. In \citet{Trager95}, the authors take surface brightness measurements from several different sources \citep{King1968,Peterson1976, Djorgovski1984, Djorgovski1985, Djorgovski1986, Djorgovski1986b} and fit a Chebyshev polynomial to the combined dataset. In the inner regions of the cluster the fit is provided in logarithmically spaced radial bins, while in the outer regions the fit is provided in linearly spaced radial bins. Therefore we treat the inner and outer surface density profiles as two separate observables, with the residual between the polynomial fit and the observational data as the uncertainty.

The surface brightness profile was converted to a luminosity profile assuming a distance of 1.85 kpc \citep{Baumgardt2021} and then converted to a surface density profile assuming the mass-to-light ratio profile of M4 determined by \citet{Baumgardt2017a}. Alternatively it is possible to include the mass-to-light ratio as a free parameter in the M2M fitting formalism \citep{deLorenzi08, Bovy18}. To account for completeness, the surface density profile was then scaled such that the total mass within the cluster's projected half-mass radius (assumed to be equal to 3/4 of the cluster's half-mass radius as determined by \citet{Baumgardt2017b}) equals  half the total cluster mass as estimated by \citet{Baumgardt2017a}.

Kinematic information also exists for M4, with the cluster's 1-D proper motion dispersion profile \citep{Vasiliev2021} and line of sight velocity dispersion profile \citep{Baumgardt2017a, Baumgardt2018, Dalgleish2020} provided within the \citet{Baumgardt2018} Galactic Globular Cluster catalogue. Uncertainties are also provided. Given the apparent isotropy of M4, it can be assumed that the velocity dispersion profiles are equal to the mean-squared velocity profiles. It is important to note that despite the kinematic information being measured over different radial ranges than the cluster's density profile, the flexibility of the M2M method allows us to incorporate data from different sources, with different coverage, and with different uncertainties.

Following the method outlined in \citet{Bovy18} to include uncertainty estimates for the M2M model cluster's properties, we generate 10 unique surface density profiles, mean-squared proper motion velocity profiles, and mean-squared line of sight velocity profiles by assuming the uncertainty in each observed datapoint is Gaussian. The M2M method is then applied to each set of unique profiles. When using the M2M method to fit observations of M4, our initial model cluster was assumed to have a density profile equal to the best fit lowered isothermal model to M4 found by \citet{deboer18}. The profile was scaled such that the cluster's total mass and effective radius are equal to the values quoted in the \citet{Baumgardt2018} Globular Cluster Catalogue. The positions and velocities of the initial model particles were again generated from that profile using LIMEPY \citep{Gieles2015}. Since M4 is denser than our mock clusters, the cluster's core evolves over a shorter timescale. We find that $\epsilon$ has to be increased from 0.005 to 10.0 in order to increase the rate at which particle masses are increased per time step. We also set $\mu$ equal to 0.01 and the smoothing parameter $\alpha$ to 0.001. Without these changes it was not possible to keep the high density core in equilibrium, which results in the model cluster's kinematics always being cooler than observed in M4.

\begin{figure*}
    \includegraphics[width=\textwidth]{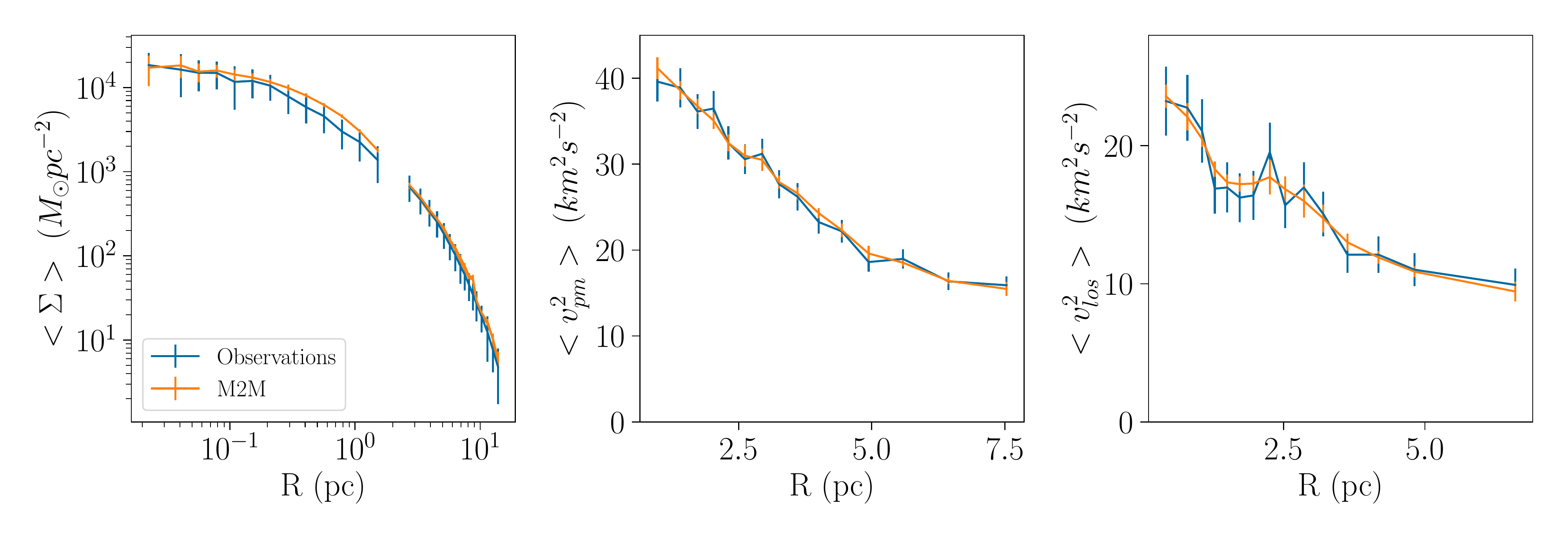}
    \caption{Surface density profile (left panel), mean-squared proper motion velocity profile (middle panel), and mean-squared line of sight velocity profile (right panel) for M4 (blue) and the average profile from 10 M2M model clusters (orange). Fitting against the cluster's projected density, proper motion velocity, and radial velocity profiles allows for the M2M method to almost exactly reproduce observations of M4.}
    \label{fig:m4}
\end{figure*}

To compare the results of fitting M2M models to random realizations of the observational data, we find the average M2M model fit to all three components and take the standard deviation of the 10 M2M models to be the uncertainty in the fit. Figure \ref{fig:m4} illustrates both the observational data and the average M2M model. The M2M method is able to reproduce the observed data very well, with a M2M model fit directly to the observations yielding an accurate representation of M4 that we have made publicly available \footnote{https://github.com/webbjj/m2mcluster/blob/main/examples/m4model.csv}. 


\section{Discussion}\label{s_discussion}

\begin{figure}
    \includegraphics[width=0.48\textwidth]{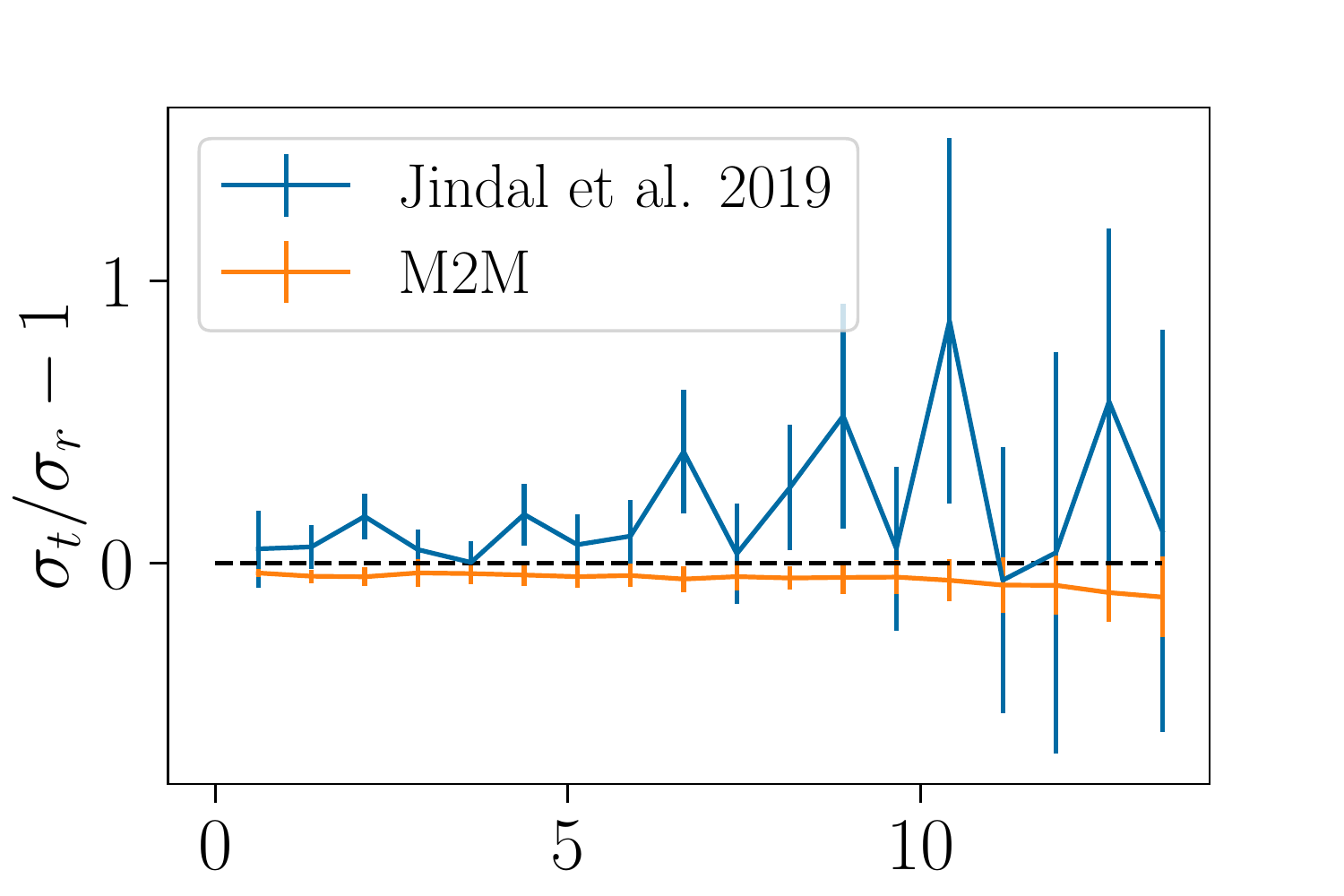}
    \caption{Observed anisotropy profile of M4 (blue) taken from \citet{Jindal2019} and average profile from 10 M2M model clusters (orange). \citet{Jindal2019} use a different dataset (\gaia\ DR2) and different membership criteria in their measurements that in this work, providing an observational constraint that is independent of our fitting process. Both the observational anisotropy profile and the M2M model clusters are consistent with being isotropic within uncertainty.}
    \label{fig:beta_plot}
\end{figure}

In the first ever application of the M2M method for modelling dynamical systems to globular clusters, we have generated a representation of the Galactic globular cluster M4 that successfully reproduces the cluster's observed surface density profile, mean-squared proper motion velocity profile, and mean-squared line of sight velocity profile. Since the method does not require that the functional form of the cluster's distribution function be assumed, the representation must be analysed in order to extract further information about the cluster. The 10 M2M model clusters have a average total mass of $\gcmass \pm \gcemass \times 10^4 M_{\odot}$, where the uncertainty is the standard deviation in the masses. The percent difference between our estimate and the value quoted in the \citet{Baumgardt2018} Galactic Globular Cluster catalogue is $6\%$, comparable to our quoted uncertainty. Furthermore, the M2M model clusters have a average a half-mass radius of \gcrad $\pm$ \gcerad pc, which has a percent difference of $2\%$ compared to the \citet{Baumgardt2018} value that is less than our uncertainty.
With a complete realization of M4 that satisfies observations of its surface density profile, mean-squared proper motion velocity profile, and mean-squared line of sight velocity profile, it becomes possible to analyze the cluster even further to determine what additional properties M4 may have given these observational constraints. For example, in Figure \ref{fig:beta_plot} we consider the cluster's anisotropy profile $\beta(r)=\sigma_t(r)/\sigma_r(r)-1$ as measured using the radial and tangential components of each model particle's proper motion. For comparison purposes, we also show the anisotropy profile measured by \citet{Jindal2019} using data from \gaia\ DR2 \citep{Gaia16a,gaia18}. It is important to note that the data used by \citet{Jindal2019} is different than the kinematics we fit our models to \citep{Baumgardt2017a,Baumgardt2018,Dalgleish2020, Vasiliev2021}, as are the criteria for determining cluster membership. Hence comparing to \citet{Jindal2019} serves as an independent check of how well the M2M model reproduces the observed properties of M4.

Comparing the independent set of anisotropy measurements from \citet{Jindal2019} to the average profile of our M2M models, we see that both cases are consistent with M4 being isotropic to within observational uncertainty. The observations, however, have significantly large uncertainties beyond 5 pc due to the lower number of stars in the dataset \citep{Jindal2019}. Also at larger distances, beyond where the M2M model is being constrained by observations, it appears the M2M model clusters are starting to become radially anisotropic. This result is not surprising given that the M2M model clusters are evolved in isolation. Including a tidal field in the simulation would ensure these stars are tidally stripped.

\begin{figure*}
    \includegraphics[width=\textwidth]{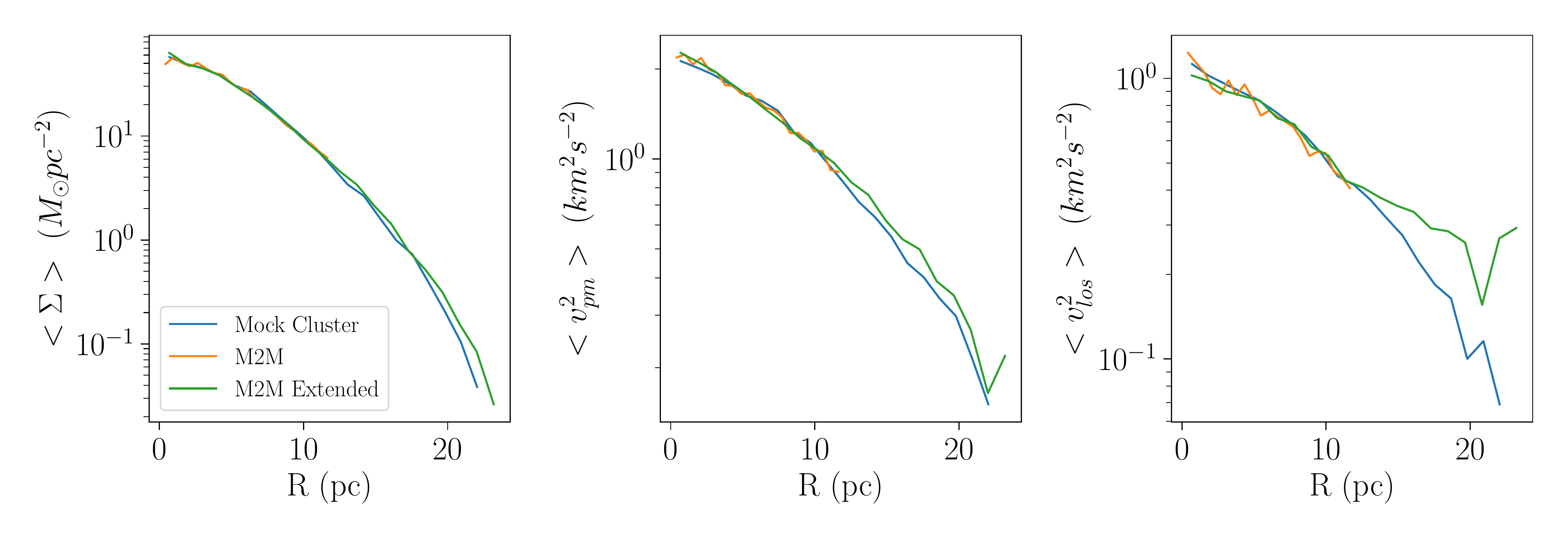}
    \caption{Surface density profile (left panel), mean-squared proper motion velocity profile (middle panel), and mean-squared line of sight velocity profile (right panel) for the isotropic mock cluster (blue) and a M2M model cluster that has only been fit against the inner 14 pc of the mock cluster (orange). While there are no observational constraints for isotropic cluster beyond approximately 12 pc, the M2M model cluster still accurately predicts how the cluster's density and kinematic profiles should extend out to the outer regions}
    \label{fig:iso_extend}
\end{figure*}

While the observational constraints on M4 only extend out to 14 pc, the cluster's tidal radius is 53.43 pc \citep{Webb2013, Baumgardt2018}. The M2M representation also allows for predictions to be made of how the cluster's spatial and kinematic properties extent out to this maximum radius. As an example, let's consider the isotropic mock cluster from Section \ref{s_results_iso}, which extends out to about 24 pc. If observations of the cluster's surface density profile, mean-squared proper motion velocity profile, and mean-squared line of sight velocity profile were only made for the inner half of the cluster, the M2M fitting method would only guarantee that the M2M model cluster matches the mock clusters within 12 pc. However as Figure \ref{fig:iso_extend} illustrates, the properties of the M2M model cluster beyond 12 pc are still closely aligned with the mock isotropic cluster. The surface density profiles in particular are in excellent agreement despite only the inner half of the cluster being used to constrain the M2M model. This strong agreement is driven by the facts that 1) stars with eccentric orbits that are in the outer regions at the final time step still passed within the inner half of the cluster and periodically have their weights adjusted and 2) outer stars are still dynamically affected by gravitational interactions with inner stars. The only discrepancy between the M2M model cluster and the isotropic mock cluster beyond the constrained region is a slight increase in the velocity dispersion at large radii in the M2M model cluster. These higher velocity stars all have eccentric orbits, with many having an eccentricity greater than 0.9. Hence they spend little time within the region constrained by the M2M algorithm and will rarely have their weights adjusted. In the presence of an external tidal field these particles would likely be stripped from the cluster.

\begin{figure*}
    \includegraphics[width=\textwidth]{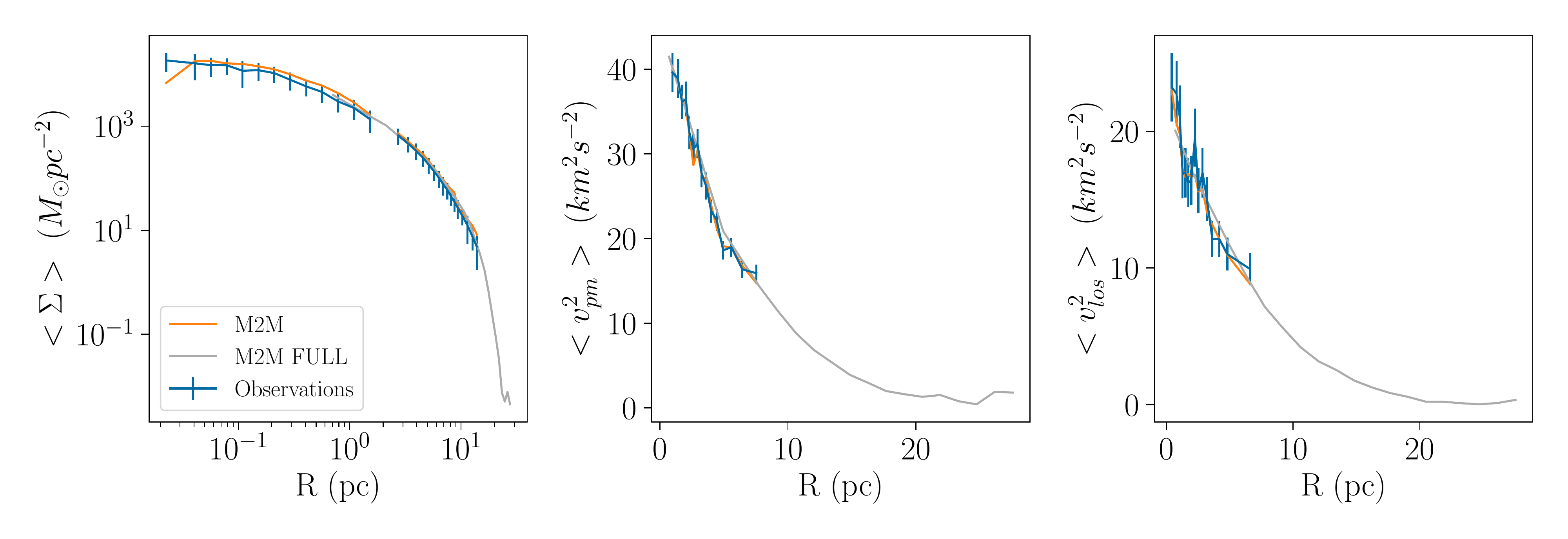}
    \caption{Same as Figure \ref{fig:m4}, except the M2M model cluster profiles are measured out to the cluster's tidal radius. While there are no observational constraints for M4 beyond approximately 14 pc, the M2M method offers predictions for how the cluster's density and kinematic profiles should extend out to the outer regions}
    \label{fig:m4_full}
\end{figure*}

Applying a similar analysis to M4, Figure \ref{fig:m4_full} again shows the surface density profile, mean-squared proper motion velocity profile, and mean-squared line of sight velocity profile for M4 and the M2M model cluster, but now extends the M2M cluster's profile out to the tidal radius. While the particles beyond the limit of the observations are not included in the comparison to the observations at the final time step, most of these particles have spent time within the inner regions of the cluster in previous time steps and are dynamically affected by the inner region particle as well. The M2M representation predicts that the cluster's surface density profile, mean-squared proper motion profile, and mean-squared line of sight velocity profile will continue to smoothly decline out to the tidal radius. 


Similar to lowered isothermal models that include potential escapers \citep{Claydon2019}, the M2M method does not require particles to be gravitationally bound to the cluster. Model stars can become energetically unbound in the simulation and will either be recaptured or pass beyond the set maximum radius (which we set to the cluster's tidal radius). However in our M2M representation, all stars in the cluster are energetically bound in the final time step.

\section{Conclusion}\label{s_conclusion}

We have demonstrated the applicability of the M2M method for modelling dynamical systems to globular clusters by fitting mock clusters generated from known distribution functions. The three-dimensional density profiles of the mock clusters are easily reproduced, regardless of the observables used in the M2M algorithm. Only when considering the outermost regions of the cluster do we find any discrepancies between the mock cluster and M2M model cluster. When fitting against the mock cluster's three dimensional density profile, density weighted mean velocity profile, or the mock cluster's density profile with all three independent velocity components, the discrepancy is negligible. When fitting against the mock cluster's projected density profile, projected density weighted mean line-of-sight velocity profile, or the mock cluster's projected density profile with two projected velocity components, the M2M model slightly overestimates the cluster's density in the outer regions.

When comparing the M2M model to the kinematic properties of the mock clusters, the inner regions are again well reproduced regardless of the observables that are being fitted against. For isotropic clusters, only fitting against the cluster's three dimensional density profile results in the M2M method overestimating the cluster's mean-squared velocity in the outer regions. However, this discrepancy is remedied when fitting against the mock cluster's density weighted mean velocity profile or the mock cluster's density profile with all three independent velocity components. Fitting against the mock cluster's projected properties, while accurately reproducing the cluster's surface density profile and projected kinematic profiles, also overestimates the cluster's mean-squared velocity in the outer regions.

For an anisotropic cluster, the M2M method is only able to reproduce the mock cluster's true anisotropy profile when fitting against its density profile with all three independent velocity components. Fitting against just the mock cluster's density profile or density weighted mean velocity profile (in three dimensions or projection) always results in an isotropic model cluster. Fitting against the mock cluster's surface density profile and projected kinematic profiles results in a M2M model that is anisotropic, but not to the same degree of the mock data.

Having demonstrated that the M2M method can accurately model star clusters, we apply the method to observations of the Galactic globular cluster M4. More specifically we apply the M2M method to measurements of M4's surface density profile, mean-squared proper motion velocity profile, and mean-squared line of sight velocity profile. The M2M method is extremely successful in generating a model star cluster that reproduce the observed properties of M4. Further study of the representation allows for the properties of the cluster's outer regions to be inferred, as well as its mass ($\gcmass \pm \gcemass \times 10^4 M_{\odot}$) and a half-mass radius (\gcrad $\pm$ \gcerad pc). 

Using the M2M method to model star clusters allows for models to be generated that do not require a functional form of the distribution to be assumed. Furthermore, the method is very flexible in terms of incorporating observations of different properties from different sources over different radial ranges and with different uncertainties. Given that we have linked our implementation of the M2M method to AMUSE \citep{PortegiesZwart2018}, there are several additional features that can be incorporated into our approach. Future applications of the M2M method to Galactic globular clusters can include a collisional $N$-body integrator, an external Milky Way-like tidal field as a free parameter, fit against the cluster's mass function to study mass segregation, and include particles that gravitationally interact within the model cluster but are too faint to be included in a surface density profile or velocity dispersion profile measurement. This latter feature would allow for estimates to be made of how massive a cluster's dark remnant population is or whether or not it contains dark matter.

\section*{Acknowledgements}

JW would like to thank Mark Gieles for helpful discussions and feedback regarding the LIMEPY models. JW would also like to thank Holger Baumgardt for sharing the M/L ratio profiles shown in the \citet{Baumgardt2018} Database and Eugene Vasiliev for helpful discussions regarding the proper motion profile of M4 measured with Gaia EDR3 in \citep{Vasiliev2021}. JB acknowledges financial support from NSERC (funding reference number RGPIN-2020-04712).

\section*{Data Availability}
 The \texttt{m2mcluster} python package is available at https://github.com/webbjj/m2mcluster. The simulated data underlying this article is available as an example within \texttt{m2mcluster}.

\bibliographystyle{mnras}
\bibliography{ref2}

\bsp

\label{lastpage}

\end{document}